\documentclass[conference]{IEEEtran}
\IEEEoverridecommandlockouts
\usepackage{cite}
\usepackage[colorlinks]{hyperref}
\hypersetup{citecolor=blue}
\usepackage{amsmath,amssymb,amsfonts}
\usepackage{algorithmic}
\usepackage{algorithm}      
\usepackage{graphicx}
\usepackage{subcaption}  
\usepackage{placeins} 
\usepackage{textcomp}
\usepackage{float}  
\usepackage{booktabs}
\usepackage{xcolor}
\usepackage{balance}
\def\BibTeX{{\rm B\kern-.05em{\sc i\kern-.025em b}\kern-.08em
		T\kern-.1667em\lower.7ex\hbox{E}\kern-.125emX}}
\begin{document}
	
	\title{Fluid Antenna-Enhanced Flexible Beamforming}
\author{Jingyuan Xu, Zhentian Zhang, Jian Dang, Hao Jiang, Zaichen Zhang
	\thanks{ }
	\thanks{Jingyuan Xu, Zhentian Zhang, Zaichen Zhang are with the National Mobile Communications Research Laboratory, Frontiers Science Center for Mobile Information Communication and Security, Southeast University, Nanjing, 210096, China. Zaichen Zhang are also with the Purple Mountain Laboratories, Nanjing 211111, China. (e-mails: \{213223473, zhangzhentian, zczhang\}@seu.edu.cn).
		
		Jian Dang is with the National Mobile Communications Research Labo-ratory, Frontiers Science Center for Mobile Information Communication and Security, Southeast University, Nanjing 211189, China, also with the Key Laboratory of Intelligent Support Technology for Complex Environments, Ministry of Education, Nanjing University of Information Science and Tech-nology, Nanjing 210044, China, and also with Purple Mountain Laboratories, Nanjing 211111, China.(email: dangjian@seu.edu.cn). 
		
		Hao Jiang is with School of Artificial Intelligence, Nanjing University of Information Science and Technology, Nanjing 210044, China. (email: jianghao@nuist.edu.cn)}
	\thanks{This work of Jian Dang and Zaichen Zhang is partly supported by the Fundamental Research Funds for the Central Universities (2242022k60001), Basic Research Program of Jiangsu (No. BK20252003), the Key Laboratory of Intelligent Support Technology for Complex Environments, Ministry of Education, Nanjing University of Information Science and Technology (No. B2202402). The work of Hao Jiang is partly supported in part by the National Natural Science Foundation of China (NSFC) projects (No. 62471238).}
}
	
	\maketitle
	
	\begin{abstract}
		Fluid antenna systems encompass a broad class of reconfigurable antenna technologies that offer substantial spatial diversity for various optimization objectives and communication tasks. Their capability to enhance spatial resolution within a fixed physical aperture makes fluid antennas particularly attractive for next-generation wireless deployments. In this work, we focus on the beamforming problem using a two-dimensional planar fluid antenna array. Since both narrow-beam and broad-beam patterns are essential in practical communication networks, enabling flexible beamforming through fluid antennas becomes an important and interesting research direction.
		We establish a unified and flexible framework that connects arbitrary beam-pattern synthesis with fluid-antenna port selection. The resulting formulation transforms beam-pattern reconstruction into a sparse regression problem, which is addressed using a tailored compressive sensing algorithm designed to operate efficiently with the fast Fourier transform (FFT). Furthermore, to ensure physically consistent phase modeling in the desired beam, we introduce an iterative FFT-based phase retrieval method. Owing to its structure, the proposed phase-refinement procedure exhibits low computational complexity and rapid convergence, requiring only one FFT and one inverse FFT per iteration.
		Simulation results demonstrate the effectiveness of the proposed flexible beamforming framework. Compared with conventional fixed-array architectures, fluid antennas exhibit significantly improved beam-pattern reconstruction accuracy, highlighting their potential for high-resolution and adaptive beamforming in future wireless systems.
		
	\end{abstract}
	
	
	\section{Introduction}
Fluid antenna systems (FASs) \cite{fas_tur,fbl_fas} refer to a class of emerging reconfigurable technologies that offer substantial spatial diversity gains by manipulating channel conditions. In general, a fluid antenna \cite{fas_pixel,fas_tur2} denotes any software-controllable fluidic, conductive, or dielectric structure capable of dynamically adjusting its radiation characteristics according to system requirements \cite{fas_1,fas_2,fas_4}. The flexibility of FAS has demonstrated significantly enhanced spatial resolution \cite{fas_5}. This capability to improve beam resolution within a fixed antenna aperture by shifting antenna structures has opened up opportunities for a wide range of practical applications, including beamforming \cite{fas_6,fas_5_5}, joint sensing\&communication \cite{fas_7,fas_8}, random access \cite{fas_3,fas_ma2}, signal processing \cite{fas_signal}, and more.

\begin{figure}[!t]
	\centering
	\includegraphics[width=\columnwidth]{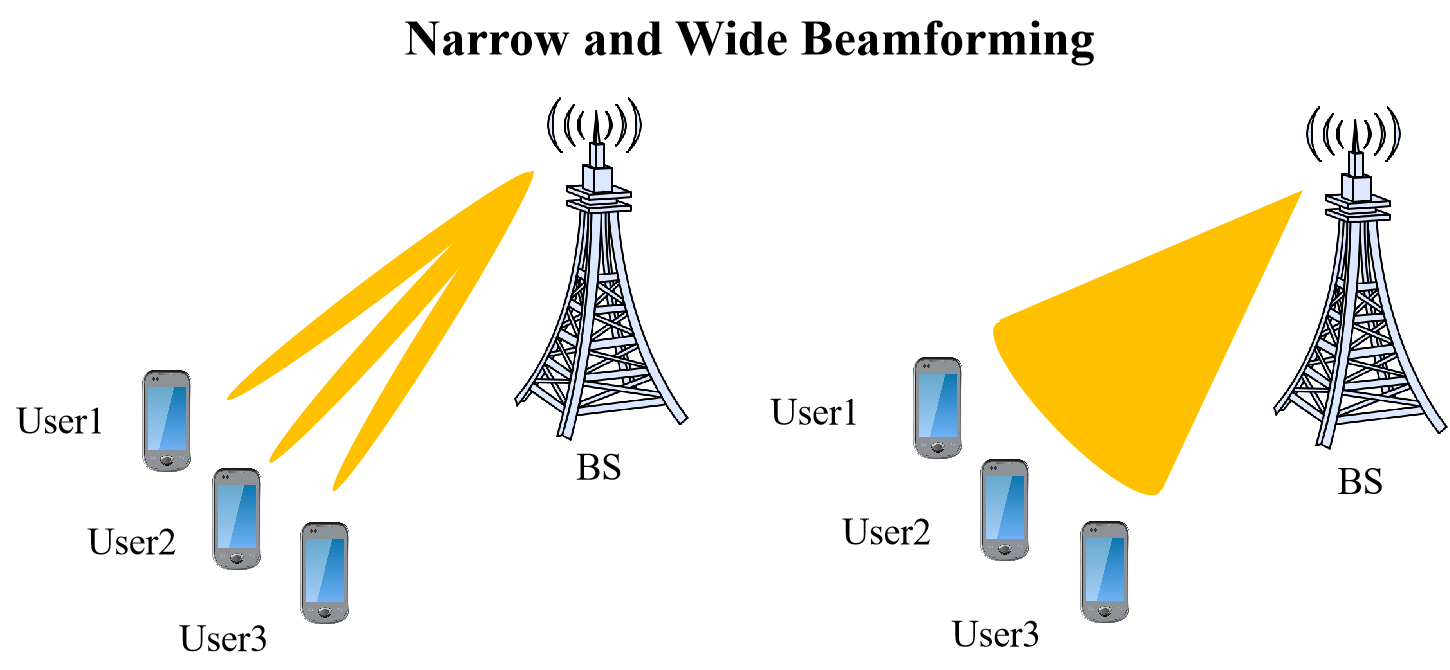}
	\caption{Illustrations of wide and narrow beams.}
	\label{fig:Wide}
\end{figure}

Particularly, this work focuses on the study of beamforming techniques in FASs. Without loss of generality, as illustrated in Fig.~\ref{fig:Wide}, we classify beamforming into narrow beam-oriented \cite{beam1} and broad beam-oriented \cite{beam2} approaches. Generally, beamforming can be characterized by noting that, based on constructive and destructive interference properties of phase-modulated sine-waves, closed-form solutions can be derived for two antennas targeting specific desired and null directions. By arranging multiple such antennas in suitable geometric patterns, beams can be synthesized to serve multiple desired directions and nulls, with the main-lobe direction potentially achieving full array gain.

Notably, neither narrow nor broad beamforming is once-and-for-all solution; nevertheless, both are indispensable. {\em User-specific (narrow) beams} provide high pointing resolution along with strong anti-interference and physical-layer security benefits, but they are sensitive to dynamic, complex environments and require substantial channel-state information. {\em Broad beamforming}, on the other hand, is crucial for synchronization, initial access, and mobility signaling, including control and broadcast channels such as the physical downlink control channel, physical broadcast channel, and synchronization signal block sweeping. These channels convey broadcast information with low per-user bitrate and thus rely on broad beams for uniform macro-sector coverage.

Motivated by these considerations, this work seeks feasible strategies for flexible beamforming in fluid antenna-based system designs. {\em Our contributions lie primarily in establishing flexible beamforming for FASs with desired beam patterns, enabling both narrow and broad beam generation in accordance with system requirements.} The main contributions are summarized as follows:
\begin{itemize}
	\item We establish a bridge between arbitrary-shape beam pattern generation and FASs. In particular, the additional spatial diversity offered by a two-dimensional plane fluid antenna allows reformulating the beamforming task as a sparse regression problem. A modified compressive sensing algorithm tailored to the fast Fourier transform (FFT) is proposed to achieve effective port selection, significantly improving beam pattern construction accuracy.
	\item To ensure reliable construction in the phase domain, we further propose an iterative FFT-based phase retrieval method that guarantees effective phase-domain modeling while avoiding destructive interference caused by phase mismatches. The proposed method is shown to be feasible in terms of both convergence and computational complexity, considering only one-time FFT and its inversion are required within single iteration.
\end{itemize}

{\em The reproducible simulation code is available at: https://github.com/BrooklynSEUPHD/Fluid-Antenna-Enhanced-Flexible-Beamforming.git.}

The remainder of this work is organized as follows: Sec.~\ref{sec.system} describes the system model and 2-D fluid antenna configurations. Sec.~\ref{sec.PSPO} presents the proposed port selection and phase retrieval methods. Numerical results are provided in Sec.~\ref{sec.simulation}. Finally, conclusions are drawn in Sec.~\ref{sec.conclusion}. {\em Notations:} Matrices and vectors are denoted by upper- and lower-case boldface letters, e.g., matrix $\mathbf{A}$ and vector $\mathbf{a}$. Let $\|\cdot\|_{2}^{2}$ denote the squared $l_2$-norm. Let $\mathbf{A}^{\mathrm{T}}$ denote the matrix transpose, and $\mathbf{A}^{\dagger}$ denote the pseudo-inverse. Moreover, $\mathcal{F}\{\cdot\}$ and $\mathcal{F}^{-1}\{\cdot\}$ denote the Fourier transform and its inverse. For matrices $\mathbf{A}$ and $\mathbf{B}$, $\langle \mathbf{A}, \mathbf{B} \rangle$ denotes their inner product. For a given matrix $\mathbf{A}$, $|\mathbf{A}|$ and $\angle\mathbf{A}$ denote its magnitude and phase responses, respectively. For a given vector $\mathbf{a}$, $|\mathbf{a}|$ denotes its Euclidean norm. The element in the $m$-th row and $n$-th column of matrix $\mathbf{A}$ is denoted by $\mathbf{A}(m,n)$.
\section{system Model}\label{sec.system}
This study considers a downlink MISO system in which each array at the base station, composed of $L$ ports in total where only $S$ fluid antennas will be selected and activated, serving user with conventionally fixed single antenna. Specifically, the base station antenna is modeled as a 2-D rectangular array with $L = M\times N$ selectable antenna positions, referred to as {\em ports}, where only $S$ ports are selected and activated. The index set of the activated antenna ports is denoted by $\mathcal{S}$, with $|\mathcal{S}|=S$. The signal is assumed to be narrowband to avoid frequency selectivity. The channel is modeled under high-frequency, slow-fading conditions, where reflected paths are neglected and only the line-of-sight component is considered. It is further assumed that the channel remains static over a given communication interval. Therefore, when a user moves from the center to the edge of a wide beam region, the base station has sufficient time to compute, infer, and adjust the beam direction.

To simplify antenna-position selection and facilitate algorithmic optimization, the antenna aperture is discretized into an $M\times N$ rectangular grid, transforming the problem into a discrete optimization over antenna positions. Moreover, because this work leverages predefined desired beam patterns for optimizing antenna selection and beamforming, the angular domain is similarly discretized, i.e., the azimuth angle $\varphi$ in the range $[-\pi/2,\pi/2]$ is discretized into $P$ points, and the elevation angle $\theta$ in the range $[-\pi/2,\pi/2]$ is discretized into $Q$ points according to the required resolution. This allows the desired beam $\mathbf{g}\in\mathbb{C}^{Z\times1}$ and the antenna weight vector ${\mathbf{w}}\in\mathbb{C}^{L\times1}$ (containing both magnitude and phase), with $Z=P\times Q$, to be written as
\begin{align}
	\underbrace{\mathbf{g} = \left[
		\begin{array}{c}
			g_1e^{-j\Phi_1} \\
			g_2e^{-j\Phi_2} \\
			g_3e^{-j\Phi_3}\\
			\vdots\\
			g_Ze^{-j\Phi_Z}
		\end{array}
		\right]}_{\text{Desired Beamforming Pattern}},
	\qquad
	\underbrace{\mathbf{w} = \left[
		\begin{array}{c}
			w_1e^{-j\Theta_1} \\
			w_2e^{-j\Theta_2} \\
			w_3e^{-j\Theta_3}\\
			\vdots\\
			w_Le^{-j\Theta_L}
		\end{array}
		\right]}_{\text{Coefficients to be Determined}},
\end{align}
where $\mathbf{G}\in\mathbb{C}^{P\times Q}$ denotes the matricized form of $\mathbf{g}$.

\begin{figure}[!t]
	\centering
	\includegraphics[width=\columnwidth]{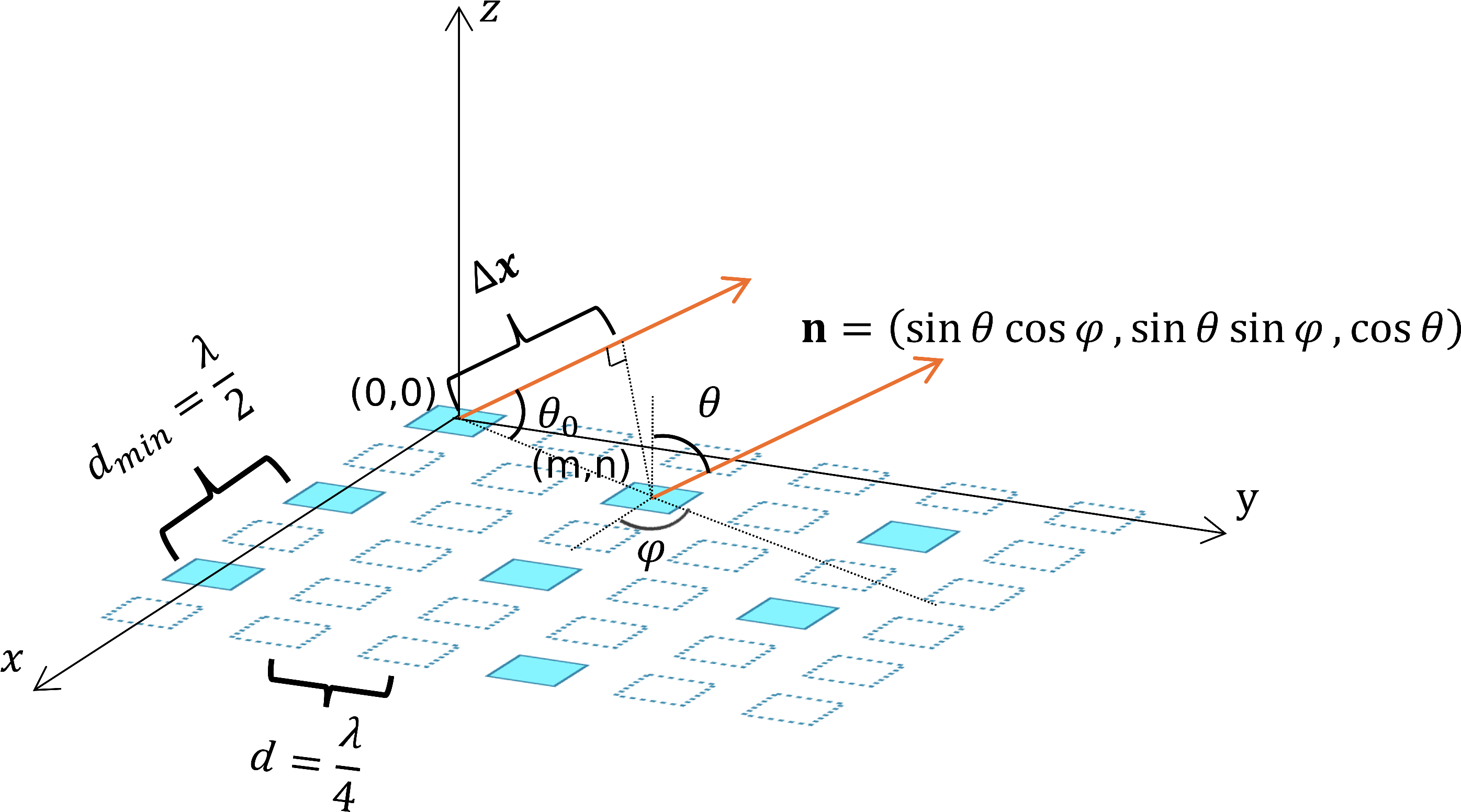}
	\caption{Illustration of the 2-D planar fluid antenna array.}
	\label{fig:array}
\end{figure}

Fig.~\ref{fig:array} illustrates the structure of the 2-D antenna array, where $d$ denotes the spacing between selectable antenna positions. To mitigate coupling effects among simultaneously activated ports, a minimum distance constraint $d_{\min}$ is imposed across selected antennas. The unit vector representing a signal departing from the origin is given by $\mathbf{n} = \left[\cos\varphi\sin\theta,\, \sin\varphi\sin\theta,\, \cos\theta\right]$. Let $\mathbf{m}=\left[x_m, y_m, 0\right]$ denote the position vector of a port. The path difference corresponds to the projection of $\mathbf{m}$ onto $\mathbf{n}$. Denoting by $\theta_0$ the angle between $\mathbf{m}$ and $\mathbf{n}$, the path difference is
\[
\Delta\mathbf{x} = |\mathbf{m}|\cos\theta_0
= \left[x_m \cos\varphi\sin\theta,\; y_m \sin\varphi\sin\theta,\; 0\right].
\]
Based on $\Delta\mathbf{x}$ and the angles $(\theta,\varphi)$, a dictionary matrix $\mathbf{D}\in\mathbb{C}^{Z\times L}$ can be constructed as the steering matrix for all available antenna positions. The entry at the $p$-th row and $q$-th column of $\mathbf{D}$ is computed as
\begin{equation}
	\begin{aligned}
		\mathbf{D}(p,q)
		=
		\exp\left\{
		-j\frac{2\pi}{\lambda}
		\left(
		x_m \cos\varphi_p \sin\theta_q
		+
		y_n \sin\varphi_p \sin\theta_q
		\right)
		\right\},
	\end{aligned}
\end{equation}
where $m\in\{1,\ldots,M\},n\in\{1,\ldots,N\},p\in\{1,\ldots,P\},q\in\{1,\ldots,Q\}$.

Given the weight vector $\mathbf{w}$, the resulting beamspace signal is $\mathbf{y}=\mathbf{D}\mathbf{w}$. This paper aims to minimize the error between the synthesized beam $\mathbf{y}$ and the desired beam $\mathbf{g}$, under the constraint that only $S$ out of $L$ ports may be activated. This leads to the optimization problem
\begin{align}\label{eq:system1}
	\min\|\mathbf{g}-\mathbf{y}\|_{2}^{2}.
\end{align}

\section{Port Selection and Phase Optimization}\label{sec.PSPO}
This section clarifies the relationship among the desired beam, port selection, and phase retrieval:
\begin{itemize}
	\item To generate the desired beam pattern, both magnitude and phase are first initialized. 
	\item Based on the desired beam pattern, the fluid antenna provides additional spatial diversity to enhance beamforming accuracy, which is achieved through port selection as discussed in Sec.~\ref{AA}. 
	\item However, the initialized phase may not align well with the practical array structure. To obtain a more suitable phase distribution for the array, an FFT-enabled iterative refinement method is proposed in Sec.~\ref{sec.phase}.
\end{itemize}
Overall, port selection in fluid antennas enables improved reconstruction capability, while phase optimization provides a better approximation model and is applied as a preprocessing step before port selection.
\subsection{Fourier Transform-based Beamforming}
Fourier series allow a function to be expressed as a sum of frequency components, with a unique correspondence between the function and its components. Similarly, the synthesized beam exhibits a corresponding relationship with each antenna element.

The actual beam $\mathbf{y}$ can be written as
\begin{align}
	{\mathbf{y}(\theta,\varphi)} = \sum_{m=1}^M \sum_{n=1}^N \mathbf{w}\left(m,n\right) e^{-j\frac{2\pi}{\lambda}(x_m \sin\theta \cos\varphi + y_n \sin\theta \sin\varphi)}.\label{eq:system3}
\end{align}

According to the Two-Dimensional Discrete Fourier Transform,
\begin{align}
	F(u,v) = \sum_{x=1}^{M} \sum_{y=1}^{N} f(x,y)e^{-j2\pi\left(\frac{ux}{M} + \frac{vy}{N}\right)}.
\end{align}

Setting $u=\sin\theta \cos\varphi/\lambda$ and $v=\sin\theta \sin\varphi/\lambda$ yields
\begin{align}\label{eq:system2}
	\mathbf{w}\left(m,n\right) = \sum_{p=1}^{P} \sum_{q=1}^{Q} \mathbf{G}(p,q) e^{j\frac{2\pi}{\lambda}(ux_m + vy_n)}.
\end{align}

The expression in \eqref{eq:system2} is particularly meaningful. Once the trigonometric functions are discretized into $u$ and $v$, the beam expression aligns exactly with the two-dimensional Fourier transform, providing clear physical insight. Thus, for any given antenna coordinate, the Fourier transform offers a closed-form expression for the corresponding beamforming weight.

However, \eqref{eq:system3} also introduces limitations and can be rewritten as
\begin{align}
	{\mathbf{y}(\theta,\varphi)} = \sum_{m=1}^M \sum_{n=1}^N \mathbf{w}\left(m,n\right) e^{-j\frac{2\pi}{\lambda}\sin\theta(x_m \cos\varphi + y_n \sin\varphi)},\label{eq:system_sin}
\end{align}
where a small value of $\sin\theta$ effectively scales the phase shift. This makes the system less sensitive to variations in $\varphi$ compared with variations in $\theta$.

\begin{figure}[!t]
	\centering
	\begin{subfigure}{\columnwidth}
		\centering
		\includegraphics[width=2.8 in]{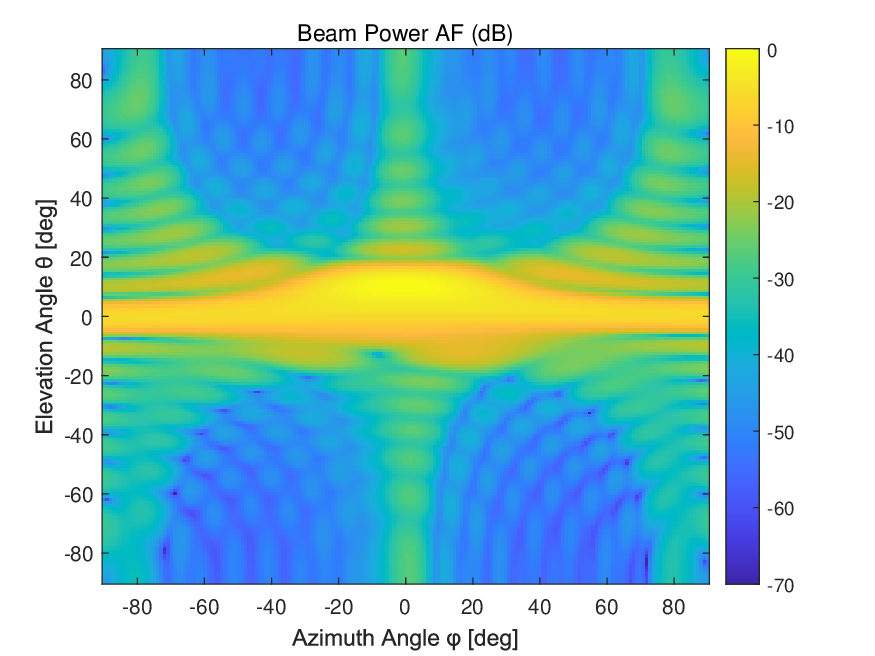}
		\caption{When $v=\sin\theta\sin\varphi/\lambda$, the beam becomes distorted near $\theta=0$.}
		\label{fig:v=sinsin}
	\end{subfigure}
	\hfill
	\begin{subfigure}{\columnwidth}
		\centering
		\includegraphics[width=2.8 in]{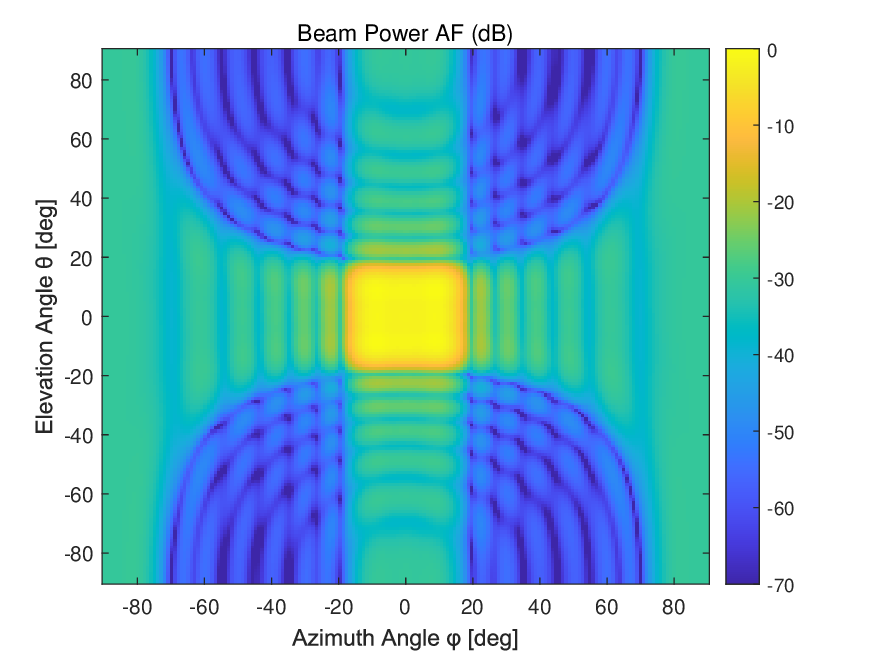}
		\caption{When $v=\sin\varphi/\lambda$, the beam is normally formed near $\theta=0$.}
		\label{fig:v=sin}
	\end{subfigure}
	\caption{Illustrations of beams near the zero elevation angle.}
	\label{fig:v}
\end{figure}

Theoretically, when $\sin\theta=0$, the beam shape becomes independent of azimuth angle $\varphi$, resulting in a long strip-like pattern, as shown in Fig.~\ref{fig:v=sinsin}. Therefore, for desired beams located in regions with small elevation angles, one may set $v=\sin\varphi$ to decouple $\varphi$ from $\theta$, ensuring that variations in $\varphi$ are not suppressed. In this work, the desired beam lies within the elevation range $[0,\pi/6]$. Hence, in subsequent simulations, $v=\sin\varphi/\lambda$ is used instead of $v=\sin\theta\sin\varphi/\lambda$, which provides more accurate reconstruction near $\sin\theta=0$, as illustrated in Fig.~\ref{fig:v=sin}.

\subsection{Port Selection Algorithm}\label{AA}
Starting from \eqref{eq:system1}, the optimization problem becomes
\begin{align}\label{eq:system4}
	\min\|\mathbf{g}-\mathbf{D}\mathbf{w}\|_{2}^{2},
\end{align}
where only $S$ of the $L$ entries in $\mathbf{w}$ are nonzero. This formulation corresponds to a sparse signal recovery problem, where $\mathbf{g}$ is the desired signal, $\mathbf{D}$ is the dictionary matrix, $\mathbf{w}$ is the sparse coefficient vector, and $S$ is the sparsity level.

The steering matrix $\mathbf{D}$ serves as a dictionary whose columns represent the beamforming contribution of each selectable antenna position. Selecting antenna ports based on a desired beam $\mathbf{g}$ resembles choosing atoms sparsely and robust compressive sensing methods \cite{OMP1,OMP2} such as orthogonal matching pursuit (OMP) have been rigorously investigated. Let the residual be initialized as $\mathbf{e}=\mathbf{g}$. Compute the inner product $\mathbf{I}=\mathbf{D}^{\mathrm{T}}\mathbf{e}$ and identify the index $i$ corresponding to the largest value, which indicates the antenna position that best matches the desired beam. Add this index to the support set $\mathcal{A}$ and compute $\mathbf{w}_{\mathcal{A}} = \mathbf{D}_{\mathcal{A}}^{\dagger}\mathbf{g}$, where $(\cdot)_{\mathcal{A}}$ denotes the data corresponding to the active index.

\begin{algorithm}[!t]
	\caption{Modified Orthogonal Matching Pursuit}
	\begin{algorithmic}[1]\label{alg.omp}
		\STATE \textbf{Input:} Matrix $\mathbf{D}$, vector $\mathbf{g}$, sparsity $S$
		\STATE \textbf{Output:} Estimate $\mathbf{w}$ solving $\min\|\mathbf{g-Dw}\|$ subject to $\|\mathbf{w}\|_0 = S$
		\STATE Normalize columns of $\mathbf{D}$
		\STATE Initialize residual $\mathbf{e}_0=\mathbf{g}$
		\STATE Initialize support set $\mathcal{A}_0=\emptyset$
		\FOR{$k=1,2,\ldots$}
		\STATE $\lambda_k=\underset{j\notin\mathcal{A}_{k-1}}{\mathrm{argmax}}\;|\langle \mathbf{D}_j,\mathbf{e}_{k-1}\rangle|$
		\STATE $\mathcal{A}_k=\mathcal{A}_{k-1}\cup\{\lambda_k\}$
		\STATE Compute beam $\mathbf{y}$ from the current support set $\mathcal{A}_k$
		\STATE Update residual $\mathbf{e}_k=\mathbf{g}-\alpha \frac{\mathbf{y}_{\mathcal{A}_k}}{|\mathbf{y}_{\mathcal{A}_k}|}$
		\ENDFOR
	\end{algorithmic}
\end{algorithm}

Traditional OMP updates the residual as $\mathbf{e}=\mathbf{g}-\mathbf{D}_{\mathcal{A}}\mathbf{w}_{\mathcal{A}}$ using least squares. However, since this work uses Fourier transform-based beamforming, the residual update must be modified. After selecting the antenna coordinates based on the inner product, the Fourier method is used to compute the beam $\mathbf{y}$. A balance coefficient $\alpha$ is introduced for residual adjustment:
\begin{align}
	\mathbf{e}=\mathbf{g}-\alpha\frac{\mathbf{y}_{\mathcal{A}}}{|\mathbf{y}_{\mathcal{A}}|}.
\end{align}

To satisfy physical array constraints, a minimum distance $d_{\min}$ between activated ports is imposed. During atom selection, both the selected position and its neighbors are excluded. These steps form the algorithm summarized in Alg.~\ref{alg.omp}.
\begin{algorithm}[!t]
	\caption{Phase Retrieval}
	\label{alg.beam}
	\begin{algorithmic}[1]
		\STATE \textbf{Input:} Desired beam $\mathbf{G}$, array size $S$
		\STATE \textbf{Output:} Refined beam pattern $\mathbf{G}$
		
		\STATE Initialize magnitude $|\mathbf{G}|$ and phase $\angle\mathbf{G}$
		
		\FOR{iterations}
		\STATE Compute spatial-domain weights $\mathbf{W}_p=\mathcal{F}^{-1}\{\mathbf{G}\}$
		\STATE Truncate $\mathbf{W}_p$ to array aperture by retaining the central block $\mathbf{W}_c\in\mathbb{C}^{\sqrt{S}\times\sqrt{S}}$
		\STATE Compute updated beam representation $\widetilde{\mathbf{G}}=\mathcal{F}\{\mathbf{W}_c\}$
		\STATE Update $\mathbf{G}\leftarrow |\mathbf{G}|\ +\angle\widetilde{\mathbf{G}}$
		\ENDFOR
		
	\end{algorithmic}
\end{algorithm}

\subsection{Phase Retrieval}\label{sec.phase}
When configuring the desired beam, prior knowledge about phase must be incorporated. Practical beams often exhibit approximately linear phase variation with respect to angle. Thus, when constructing the desired beam $\mathbf{g}$ or $\mathbf{G}$, a linear phase is added to avoid unrealistic phase values that would degrade beamforming. However, this linear initialization may not match the underlying antenna array, explaining the necessity of phase retrieval \cite{iteration}. 

To address this issue, we propose an iterative Fourier algorithm to refine the phase of the desired beam. The procedure uses two FFT operations per iteration:

\begin{itemize}
	\item Initialize magnitude and phase of $\mathbf{G}$ as $|\mathbf{G}|$ and $\angle\mathbf{G}$.
	\item Apply inverse Fourier transform to obtain $\mathbf{W}_p=\mathcal{F}^{-1}\{\mathbf{G}\}$.
	\item Tail $\mathbf{W}_p$ to the antenna array dimensions by retaining only the central block:
	\[
	\mathbf{W}_p \rightarrow
	\begin{bmatrix}
		\mathbf{W}_c & \mathbf{0} \\
		\mathbf{0} & \mathbf{0}
	\end{bmatrix},
	\]
	where $\mathbf{W}_c\in\mathbb{C}^{\sqrt{S}\times \sqrt{S}}$.
	\item Apply Fourier transform on $\mathbf{W}_c$ to obtain $\widetilde{\mathbf{G}}$ and extract updated phase.
	\item Update the desired beam as $\mathbf{G}\leftarrow |\mathbf{G}|+\angle\widetilde{\mathbf{G}}$.
	\item Repeat until convergence.
\end{itemize}

Because the initialized phase $\angle\mathbf{G}$ may not conform to the array structure, the iterative process gradually adapts it into a more realistic phase distribution that reduces destructive interference.

\begin{figure}[!t]
	\centering
	\begin{subfigure}{\columnwidth}
		\centering
		\includegraphics[width=3.05 in]{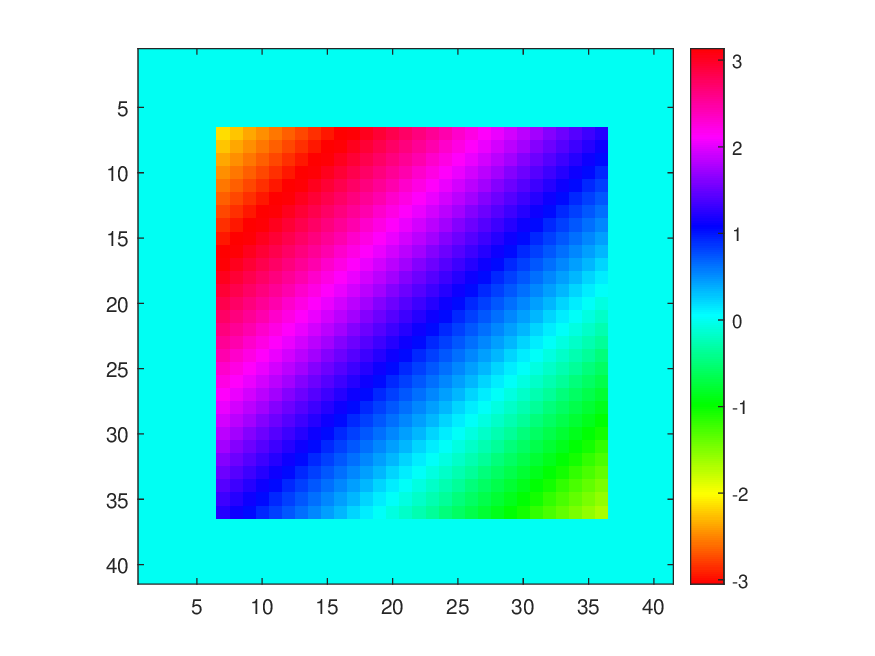}
		\caption{Linear phase with $k=0.1$}
		\label{fig:Linear Phase}
	\end{subfigure}
	\hfill
	\begin{subfigure}{\columnwidth}
		\centering
		\includegraphics[width=3.05 in]{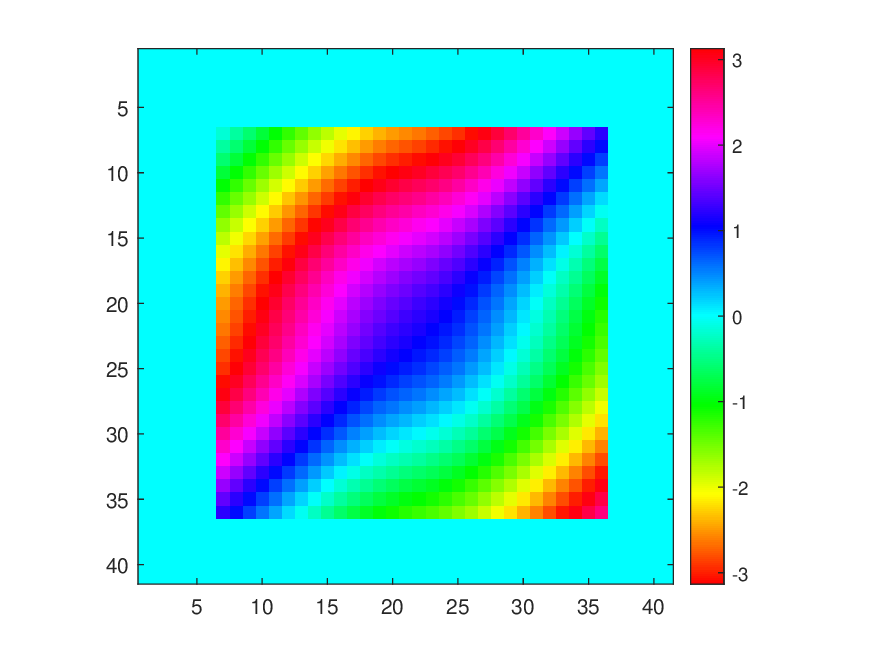}
		\caption{Optimized phase with $k=0.1$}
		\label{fig:Optimized phase}
	\end{subfigure}
	\caption{Illustrations of phase domain $\angle\mathbf{G}$ with and without phase retrieval.}
	\label{fig:Phase Distribution}
\end{figure}

An example is shown in Fig.~\ref{fig:Phase Distribution}. Fig.~\ref{fig:Linear Phase} presents the linear initialization, while Fig.~\ref{fig:Optimized phase} shows the optimized phase corresponding to the desired beam. Unlike the uniform structure of the linear phase, the optimized phase exhibits a central elevation that more closely resembles practical beam behavior. This refined phase is more suitable for beamforming, and the performance improvement will be demonstrated in simulations.

\section{Simulation}\label{sec.simulation}

This section evaluates the performance of the proposed position–phase optimization framework through far-field beamforming simulations. The inter-element spacing of the physical antenna is set to $\lambda/2$. To ensure a comparable aperture size while avoiding coupling, the interval between selectable fluid-antenna ports is configured as $\lambda/4$. A minimum separation constraint of $d_{\min}=\lambda/2$ is imposed during port selection. The desired beam is initialized with a linear phase of slope $k=0.1$, chosen based on empirical tuning. Other simulation parameters are summarized in Table~\ref{tab:my_label}.

\begin{table}[!t]
	\centering
	\caption{System Parameters}
	\begin{tabular}{ccc}
		\toprule
		\textbf{Variable} & \textbf{Symbol} & \textbf{Value}\\
		\midrule
		Azimuth angle range & $\varphi$ & $[-\pi/2,\pi/2]$\\
		Elevation angle range & $\theta$ & $[-\pi/2,\pi/2]$ \\
		Angular quantization scale & $P,Q$ & 180 \\
		Number of active antennas & $S$ & 256 \\
		Number of selectable ports & $L$ & 1024 \\
		One-dimensional array size & $L_a$ & 16 \\
		Minimum spacing & $d_{\min}$ & ${\lambda}/{2}$ \\
		Port-grid spacing & $d$ & ${\lambda}/{4}$ \\
		Initial phase slope & $k$ & 0.1 \\
		Residual update coefficient & $\alpha$ & $-0.01$ \\
		\bottomrule
	\end{tabular}
	\label{tab:my_label}
\end{table}

\begin{figure}[!t]
	\centering
	\includegraphics[width=2.85 in]{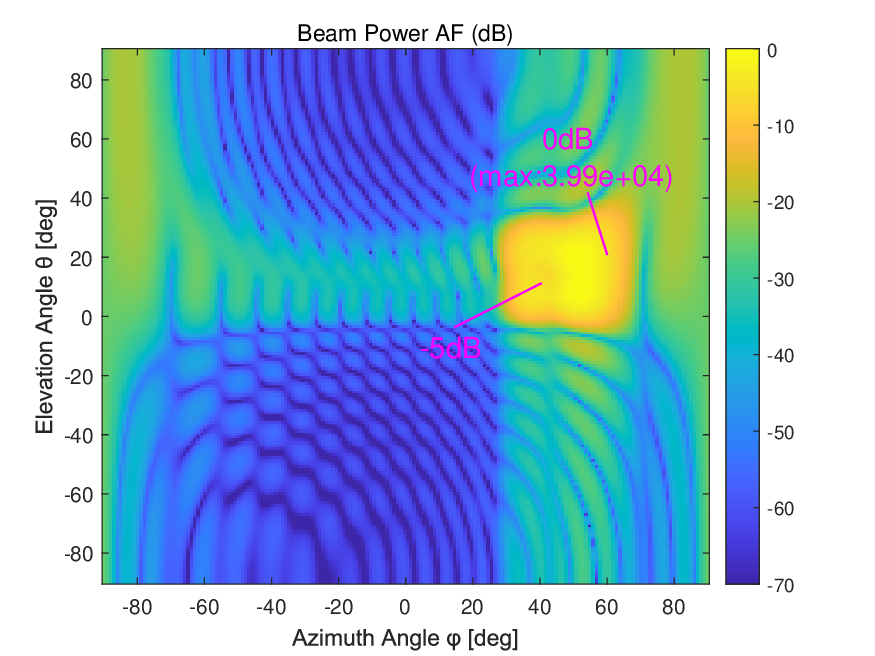}
	\caption{Normalized beam heatmap for a fixed antenna array, showing structured and predictable side-lobe patterns.}
	\label{fig:fixed}
\end{figure}

\begin{figure}[!t]
	\centering
	\includegraphics[width=2.85 in]{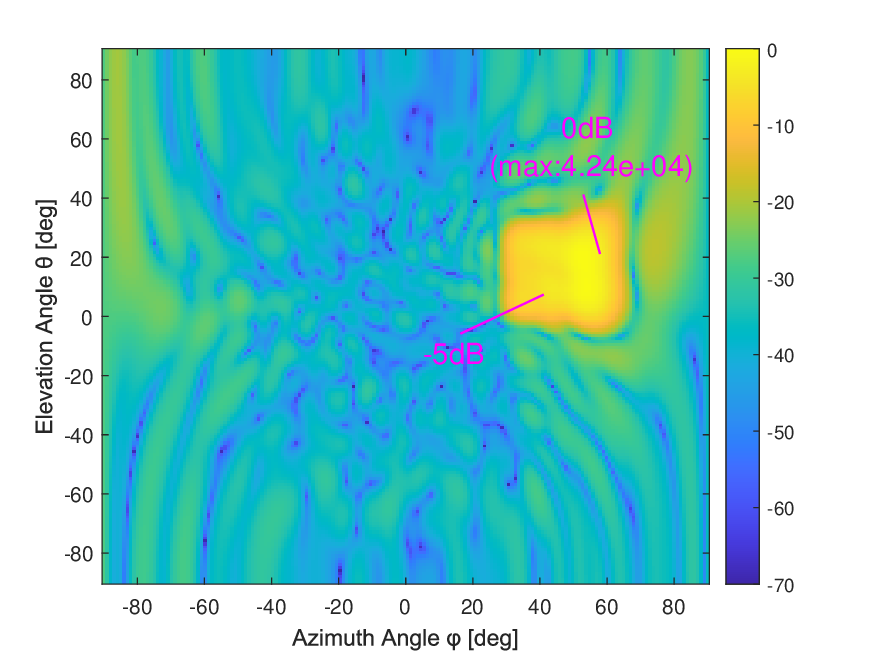}
	\caption{Normalized beam heatmap for a fluid antenna array. Randomized port selection suppresses structured side lobes, enhancing power concentration and physical secrecy.}
	\label{fig:fluid}
\end{figure}

We first visualize the beam patterns generated by different array architectures. The desired beam $\mathbf{G}$ is steered toward $\varphi_1\in[\pi/6,\pi/3]$ and $\theta_1\in[0,\pi/6]$. Fig.~\ref{fig:fixed} shows the normalized beam of a conventional fixed array, while Fig.~\ref{fig:fluid} illustrates that of the proposed fluid antenna using port optimization. The fluid-antenna beam exhibits noticeably reduced reconstruction error and significantly higher main-lobe SNR. The suppression of periodic side lobes in Fig.~\ref{fig:fixed} originates from the irregular spatial sampling associated with port selection, which effectively disrupts the deterministic radiation pattern. {\em Beyond improved beam quality, such irregular side-lobe behavior also provides an inherent advantage for physical-layer secrecy, as the resulting beam becomes harder to infer or predict.}

\begin{figure}[!t]
	\centering
	\includegraphics[width=3.2 in]{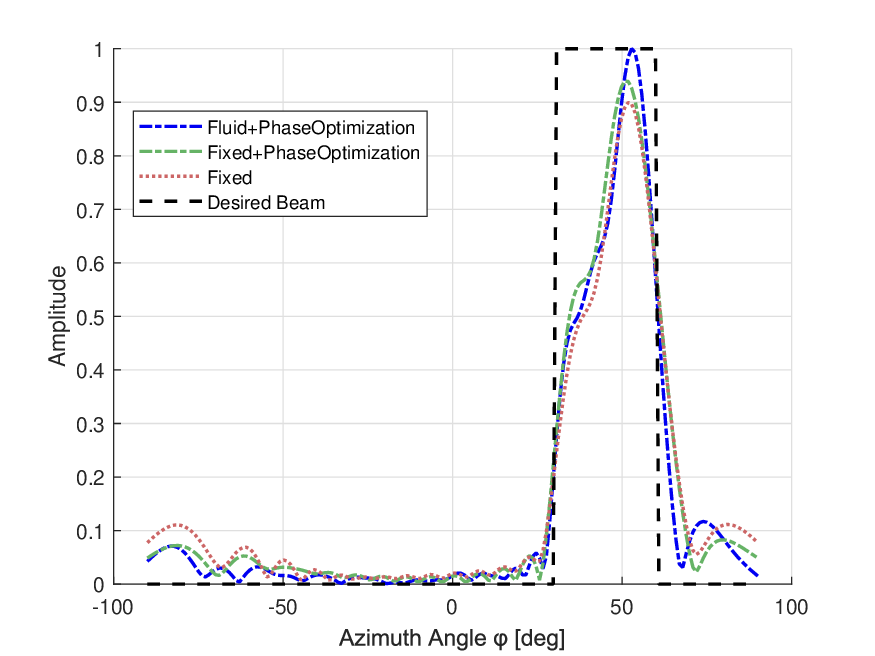}
	\caption{Beam cross-section at $\theta=20^\circ$ under different optimization schemes.}
	\label{fig:CrossSection1}
\end{figure}

\begin{figure}[!t]
	\centering
	\includegraphics[width=3.2 in]{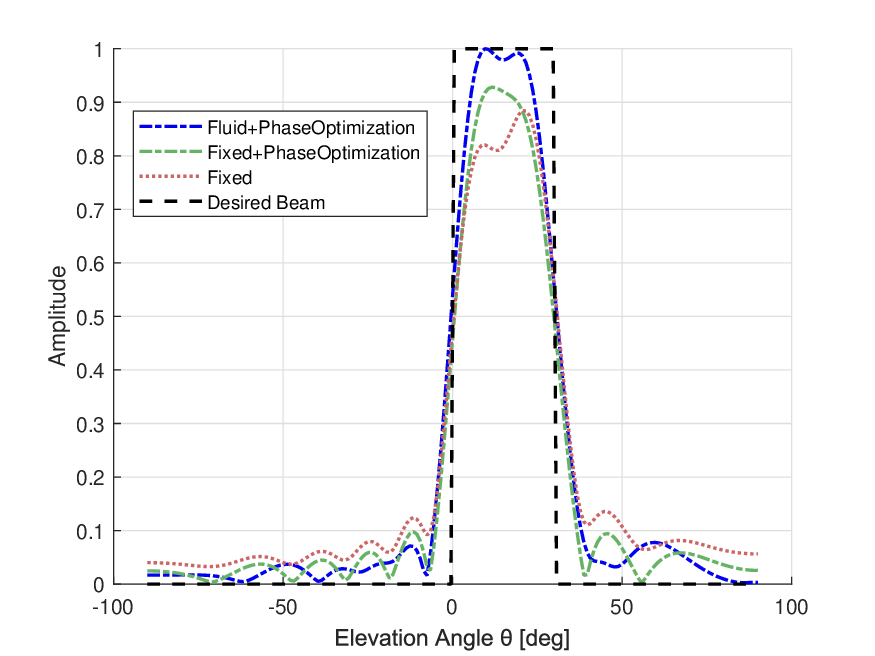}
	\caption{Beam cross-section at $\varphi=55^\circ$ under different optimization schemes.}
	\label{fig:CrossSection2}
\end{figure}

Next, we compare different algorithmic configurations. Fig.~\ref{fig:CrossSection1} and Fig.~\ref{fig:CrossSection2} show cross-sectional beam profiles at $\theta=20^\circ$ and $\varphi=55^\circ$, respectively, for three schemes:  
\begin{itemize}
	\item[-] Fixed array, in short: ``{\em Fixed}'',  
	\item[-] Fixed array with phase optimization, in short: ``{\em Fixed+PhaseOptimization}'',  
	\item[-] Fluid antenna with phase optimization, in short: ``{\em Fluid+PhaseOptimization}''.
\end{itemize}
All beams are normalized for fair comparison. The main-lobe widths across the three schemes are similar, confirming that the Fourier-domain beamforming framework provides a consistent baseline. However, the Fixed+PhaseOptimization and Fluid+PhaseOptimization curves exhibit significantly greater energy concentration, validating that both phase retrieval and port selection enhance focusing capability. Notably, the fluid-antenna configuration consistently achieves the highest main-lobe gain due to the additional spatial degrees of freedom provided by port optimization.

It is also worth emphasizing that these results correspond to a favorable initial phase estimate ($k=0.1$). Under poorer initialization, the benefit of the proposed FFT-based phase retrieval becomes more pronounced, since unoptimized schemes are substantially more sensitive to phase distortion.

	\section{Conclusion}\label{sec.conclusion}
This paper presented a flexible beamforming framework for fluid antennas that combines FFT-based phase refinement with a modified OMP port-selection algorithm. By leveraging the Fourier structure of beam patterns, the proposed method enables accurate phase adaptation and efficient position optimization under physical spacing constraints. Simulations verified that the joint approach significantly improves main-lobe power concentration, suppresses structured side lobes, and provides inherent physical-layer secrecy through irregular radiation patterns. The results further show that phase refinement is especially beneficial when initial phase conditions are imperfect. Overall, the proposed framework demonstrates the performance and robustness advantages of fluid antennas over fixed arrays, offering a promising foundation for future beamforming research.

	\balance
	
\end{document}